\documentclass{iopart}

\usepackage{graphicx}

\newcommand{\hide}[1]{}

\begin{document}

\title[Positron emitting nuclei in proton and heavy-ion therapy]{Distributions of 
positron-emitting nuclei in proton and carbon-ion therapy studied with GEANT4}
    
\author{Igor~Pshenichnov$^{1,2}$, Igor~Mishustin$^{1,3}$ and Walter~Greiner$^1$}  
\address{$^1$ Frankfurt Institute for Advanced Studies, Johann Wolfgang Goethe University, 
60438 Frankfurt am Main, Germany}
\address{$^2$ Institute for Nuclear Research, Russian Academy of Science, 117312 Moscow, Russia}
\address{$^3$ Kurchatov Institute, Russian Research Center, 123182 Moscow, Russia}

\begin{abstract}

Depth distributions of positron-emitting nuclei 
in PMMA phantoms are calculated within a Monte Carlo model for Heavy-Ion Therapy (MCHIT) 
based on the GEANT4 toolkit (version 8.0). The calculated total production rates 
of $^{11}$C, $^{10}$C and $^{15}$O nuclei are compared with experimental data and with 
corresponding results of the FLUKA and POSGEN codes. 
The distributions of e$^+$ annihilation points are obtained by simulating radioactive decay of unstable 
nuclei and transporting positrons in surrounding medium.  A finite spatial resolution of the Positron Emission
Tomography (PET) is taken into account in a simplified way.
Depth distributions of $\beta^+$-activity as seen by a PET scanner are calculated and compared 
to available data for  PMMA phantoms. The calculated  $\beta^+$-activity profiles are in 
good agreement with PET data for proton and $^{12}$C beams at energies suitable for particle therapy.     
The MCHIT capability to predict the $\beta^+$-activity and dose distributions in 
tissue-like materials of different chemical composition is demonstrated.
  
\end{abstract}

\submitto{\PMB}

\ead{pshenich@fias.uni-frankfurt.de}


\section{Introduction}

Beams of charged particles, protons and light nuclei, are proved to be very efficient  
for the radiation therapy of deep-seated solid tumours (Castro \etal 2004)\hide{~\cite{Castro:etal:2004}}.
Their selective and strong impact on a tumour is possible due to a 
maximum energy deposition at the end of their range in matter (the Bragg peak).
According to the statistical data of Particle Therapy Cooperative Group (PTCOG 2006)\hide{~\cite{PTCOG:2006}}
over 48000 patients were treated worldwide with proton or ion beams by July 2005. 

Light nuclei, e.g. carbon ions, have an additional potential advantage 
compared to protons associated with their enhanced  
relative biological effectiveness (RBE) 
(Kraft 2000\hide{~\cite{Kraft:2000}}). Treatment schemes using the carbon-ion irradiation demonstrated
good clinical results (Schulz-Ertner \etal 2004,\hide{~\cite{Schulz-Ertner:2004}}
Tsujii \etal 2004,\hide{~\cite{Tsujii:etal:2004}}  Kanai \etal 2006\hide{~\cite{Kanai:etal:2006}}).
 
In most cases so far (see PTCOG 2006)\hide{~\cite{PTCOG:2006}} the  
dedicated therapy facilities use either proton or heavy-ion beams. 
Since 2001, the Hyogo Ion Beam Medical Center (HIBMC) is 
the first facility in the world  which operates with two types 
of beams, protons and carbon-ions, in a single location (Hishikawa \etal 2004,\hide{~\cite{Hishikawa:etal:2004}} 
Mayahara \etal 2005\hide{~\cite{Mayahara:etal:2005}}). 
Several new facilities are planned or under construction in Europe:
HIT in Germany (Haberer \etal 2004,\hide{~\cite{Haberer:etal:2004}} Heeg \etal 2004\hide{~\cite{Heeg:etal:2004}}), 
CNAO in Italy (Amaldi 2004)\hide{~\cite{Amaldi:2004}}, 
ETOILE in France (Bajard \etal 2004)\hide{~\cite{Bajard:etal:2004}}
and MedAustron in Austria (Griesmayer and Auberger 2004)\hide{~\cite{Griesmayer:and:Auberger:2004}}. 
These particle therapy centres will also have proton and carbon-ion beams on their sites. 
Each of the two types of beams or their combination will be available for patient treatment.   

Successful particle therapy requires thorough treatment planning, which must
include optimisation between positive and negative effects of the beam.
A required dose of radiation has to be delivered to a tumour while sparing surrounding 
healthy tissues and organs at risk. 
The optimisation of doses requires a reliable method for calculating the beam energy deposition
in highly heterogeneous human tissues.
Up to now, mostly deterministic methods have been used by different groups for dose calculations in hadron 
therapy, see e.g. Hong \etal (1996), Kr\"{a}mer \etal (2000), 
J\"{a}kel \etal (2001)\hide{~\cite{Hong:etal:1996,Kramer:etal:2000,Jakel:etal:2001}}.
However, nowadays a Monte Carlo simulation of particle transport in human body can be used
as an alternative method for dose calculations. 
Increasing power of modern computers makes this method less restrictive in terms of the CPU time
(Rogers 2006)\hide{~\cite{Rogers:2006}}.

In view of the prospective wide use of both proton and carbon-ion therapy in several 
medical centres worldwide, one should think of employing common computational tools for proton and 
ion therapy based on the Monte Carlo approach.
We believe that the GEANT4 toolkit (Agostinelli \etal 2003)\hide{~\cite{Agostinelli:etal:2003}}
is well suited for this purpose. This toolkit was created by an international collaboration 
of physicists and programmers for 
basic research in nuclear and particle physics. This is an open-source project based on modern 
techniques of  programming and visualisation. It is capable of simulating a wide 
range of physical processes in extended media, which makes it useful for applications in medical physics.

The GEANT4 object-oriented toolkit is written in C++ and contains classes and methods which provide
the basic functions for simulations (GEANT4-Documents 2006\hide{~\cite{GEANT4-Documents:2006}}), 
e.g. handling setup geometry, tracking primary and secondary
particles, calculation of their energy loss and energy deposition in matter, run management, 
visualisation and user interface. 
On this basis, a specific application can be developed for each particular task. 
The results on validation of the GEANT4 toolkit for proton therapy 
(Jiang and Paganetti 2004)\hide{~\cite{Jiang:and:Paganetti:2004}} 
and for therapy with electrons and photons (Carrier \etal 2004,\hide{~\cite{Carrier:etal:2004}} 
Larsson \etal 2005\hide{~\cite{Larsson:etal:2005}}) have been presented recently. 
A GEANT4-based application for emission tomography (GATE) was developed by the 
Open-GATE collaboration (Jan \etal 2004)\hide{~\cite{Jan:etal:2004}} and successfully used in
Monte Carlo simulations of commercial PET scanners (Schmidtlein \etal 2006,\hide{~\cite{Schmidtlein:2006}}      
Lamare \etal 2006\hide{~\cite{Lamare:2006}}). The GEANT4 toolkit is also used for
simulations in brachytherapy (Enger \etal 2006\hide{~\cite{Enger:etal:2006}},
Perez-Calatayud \etal 2006\hide{~\cite{Perez-Calatayud:etal:2006}}).

In our recent paper (Pshenichnov \etal 2005)\hide{~\cite{Pshenichnov:2005}} we 
have made a first attempt to validate the GEANT4 toolkit of version 7.0
(GEANT4-Webpage 2006\hide{~\cite{GEANT4-Webpage:2006}})
for heavy-ion therapy simulations. We have developed a GEANT4-based application for 
Monte Carlo simulations of Heavy-Ion Therapy (MCHIT).
The depth-dose distributions in tissue-like
media, calculated within this model, are in a good agreement with experimental data. The relative 
contribution to the total dose due to secondary neutrons was quantitatively evaluated for proton and ion beams.   

In the present work we extend the applicability of the MCHIT model 
to calculations of secondary $\beta^+$-radioactivity induced by proton and heavy-ion beams in tissue-like media.
Currently, the MCHIT application is based on the version 8.0 (with patch 01) of GEANT4 toolkit.

\section{Dose and range monitoring in particle therapy by positron emission tomography}\label{Concept}

The main advantage of proton and heavy-ion therapy consists in the possibility to deliver a prescribed dose 
to a tumour volume while sparing surrounding healthy tissues and organs at risk. 
This requires thorough verification of the proton and heavy-ion ranges in patient's body.
Tissue inhomogeneities and local anatomical changes between fractions 
of particle therapy may lead to some differences between the actual dose delivered 
to a patient and the prescribed dose.

Two methods for dose verification in proton and heavy-ion therapy were proposed.  
In the first method, the beams of $\beta^+$-radioactive nuclei $^{19}$Ne, $T_{1/2}= 17.22$~s,
(Tobias \etal 1977)\hide{~\cite{Tobias:etal:1977}},  
$^{11}$C, $T_{1/2}=20.39$~min,  (Urakabe \etal 2001)\hide{~\cite{Urakabe:etal:2001}} or 
$^{10}$C, $T_{1/2}=19.255$~s, (Iseki \etal 2004)\hide{~\cite{Iseki:etal:2004}} 
are injected into the patient, fixed in the proper position, just before
the therapeutic treatment with the beams of stable nuclei, $^{20}$Ne or $^{12}$C, respectively.
However, the practical application of this method is limited by 
high costs of the radioactive beams.

More frequently another method is used which utilises the fact that  
$\beta^+$-radioactive nuclei are produced in fragmentation 
reactions taking place in tissues during irradiation with beams of protons or stable nuclei, e.g. $^{12}$C. 
In proton therapy target fragmentation reactions produce $^{15}$O,
$T_{1/2}=122.24$~s, among other fragments. As was early realized 
(Bennett \etal 1975, 1978)\hide{~\cite{Bennett:etal:1975,Bennett:etal:1978}},
this can be used for controlling the proton beam localisation in tissues. Later, a similar auto-activation technique, 
but involving $^{10}$C and $^{11}$C nuclei produced in fragmentation of $^{12}$C projectiles,
was proposed for dose monitoring in carbon-ion therapy 
(Pawelke \etal 1996, 1997)\hide{~\cite{Pawelke:etal:1996,Pawelke:etal:1997}}.
  
Unstable nuclei $^{10}$C, $^{11}$C and  $^{15}$O  
undergo $\beta^+$-decay: $A(Z,N) \rightarrow A(Z-1,N+1)+\rm{e}^+ + \nu_e$. Due to the three-body
kinematics, the energy released in the transition of a bound proton into a bound neutron 
is partly carried away also by a neutrino, so that emitted positrons have a continuous energy spectrum.
The maximum energy of positrons emitted e.g. by $^{11}$C nucleus is 960 keV with the
average energy of 386 keV. Such positrons can travel up to 4 mm in the human body before they stop and 
annihilate on surrounding electrons: ${\rm e}^+{\rm e}^-\rightarrow\gamma\gamma$. Due to multiple
scattering of positrons, their path in a tissue is far from the straight line,
and their average penetration depth is shorter than their actual path. This difference is 
described by a detour factor (Fern\'{a}ndez-Varea \etal 1996)\hide{~\cite{Fernandez-Varea:etal:1996}}, 
which is estimated, for example, for 1 MeV electrons or positrons in water as $\sim 0.5$.
As a result, most of the positrons emitted by $^{11}$C annihilate within $\sim 2$ mm 
from their emission point (see also Levin and Hoffman 1999)\hide{~\cite{Levin:and:Hoffman:1999}}.

In the course of multiple scattering the positrons slow down, so that ${\rm e}^+{\rm e}^-$
annihilation take place practically at rest. Therefore,  
the angle between the momenta of the emitted photons is close 
to $180^0$. They are registered by 
detectors outside the patient body. Corresponding reconstruction algorithms make possible to
obtain the spatial distribution of annihilation points, which should be close to the 
distribution of positron-emitting nuclei. It is further expected that the spatial distribution of 
$\beta^+$-activity induced by proton and heavy-ion 
beams is strongly correlated with the corresponding dose distribution. However, 
there exists no simple way to express the dose distribution 
in terms of the $\beta^+$-activity distribution.

In practice, the problem is solved in a few steps. First, the $\beta^+$-activity distributions
are measured in advance in experiments with phantoms for different beam energies and doses.
In particular, such measurements have been performed for proton beams in phantoms made of  
polymethylmethacrylate (PMMA), ${\rm C}_5{\rm H}_8{\rm O}_2$ or lucite ($\rho=1.18$~g/cm$^3$) 
by Oelfke \etal (1996)\hide{~\cite{Oelfke:etal:1996}} and 
Parodi \etal (2002)\hide{~\cite{Parodi:etal:2002}}. 
The $\beta^+$-activity measurements for $^{20}$Ne beams in PMMA were made 
by Enghardt \etal (1992)\hide{~\cite{Enghardt:etal:1992}}, 
and later for $^{12}$C beams by 
Pawelke \etal (1997)\hide{~\cite{Pawelke:etal:1997}},  
 P\"{o}nisch \etal (2004)\hide{~\cite{Poenisch:etal:2004}} and
Parodi (2004)\hide{~\cite{Parodi:2004}}.
 
Second, the models capable of calculating both the dose and $\beta^+$-activity distributions 
are validated with these data. For example, a phenomenological model for the proton transport
incorporating data on isotope production was used by Oelfke \etal (1996)\hide{~\cite{Oelfke:etal:1996}}  
to calculate the dose and $\beta^+$-activity distributions in homogeneous media.   
Parodi and Enghardt (2000)\hide{~\cite{Parodi:and:Enghardt:2000}},  
Parodi \etal (2002)\hide{~\cite{Parodi:etal:2002}}, 
P\"{o}nisch \etal (2004)\hide{~\cite{Poenisch:etal:2004}}
simulated the transport of protons and carbon ions 
in PMMA phantoms with FLUKA and POSGEN codes, respectively.
In the present paper we propose to use the GEANT4 toolkit for calculating 
the dose and $\beta^+$-activity distributions. 

Third, the dose applied to a patient in a therapeutic treatment can be verified via comparison of  
the $\beta^+$-activity distribution in the patient body with the distribution predicted by 
the model for the same dose.

\section{GEANT4 physics models used in MCHIT}

We use the version 8.0 (with patch 01) of the GEANT4 toolkit 
(GEANT4-Webpage 2006)\hide{~\cite{GEANT4-Webpage:2006}} to build a Monte Carlo 
model for Heavy-Ion Therapy (MCHIT). Currently this model is capable of 
calculating the three-dimensional distributions of  dose and $\beta^+$-activity in 
tissue-like media.
In the present study we use homogeneous phantoms  
represented by a water cube or by a cube made of PMMA. 
Calculations can also be made for phantoms representing bone, liver or muscle tissues.
In simulations the phantom is irradiated 
by beams of protons or ions with given beam size, emittance, angular convergence/divergence 
and energy distribution. The phantom material and size, as well as all beam parameters can be 
set via user interface commands individually for each run. 
     
According to the GEANT4 concept, the set of physical models which are relevant for a particular 
problem should be activated by the user 
via a set of commands. A detailed description of physical models included in GEANT4 is given 
in the Physics Reference Manual (GEANT4-Documents 2006)\hide{~\cite{GEANT4-Documents:2006}}. 
In order to facilitate this selection, the GEANT4 developers recommend to start either from   
available examples of previously developed applications, or from the so-called predefined physics lists.

Here we briefly describe the choice of models and main parameters used in our calculations.  
In MCHIT the energy loss of primary and secondary charged particles due to electromagnetic processes is 
described via a set of models called 'standard electromagnetic physics'. It accounts for energy loss and
straggling due to interaction with atomic electrons as well as multiple Coulomb scattering on atomic nuclei.   
 
At each simulation step, the ionisation energy loss of a charged particle is calculated according to the
Bethe-Bloch formula. The mean excitation potential of 
water molecules was set to 77 eV, i.e. to the value which better describes the set of available data on 
depth-dose distributions, both for proton and carbon-ion beams, in the range of 
therapeutic energies of
80 - 330 MeV per nucleon. For PMMA, the mean excitation potential was set to 
68.5 eV, a default value used in GEANT4 for this material.  

The bremsstrahlung processes for electrons and positrons were activated in the simulations along with the
annihilation process for positrons. For photons, Compton scattering, the conversion into an electron-positron
pair and the photoelectric effect were included.

Two kinds of hadronic interactions are considered in the MCHIT model:
(a) elastic scattering of hadrons on target protons and nuclei, which dominate 
at low projectile energies, and (b) inelastic nuclear reactions induced by 
fast hadrons and nuclei.
The model of nucleon-nucleon elastic scattering is based on a parameterisation of experimental data in
the energy range of 10-1200 MeV. At higher energies the hadron-nucleus elastic
scattering is simulated within the Glauber model (GEANT4-Documents 2006)\hide{~\cite{GEANT4-Documents:2006}}.  

Overall probability of hadronic interactions for nucleons and nuclei propagating in the 
media is determined by the total inelastic cross section for nucleon-nucleus and nucleus-nucleus collisions.
Parameterisations by Wellisch and Axen (1996)\hide{~\cite{Wellisch:and:Axen:1996}} 
that best fit experimental data were used  to describe the total reaction
cross sections in nucleon-nucleus collisions. 
Systematics by Shen \etal (1989)\hide{~\cite{Shen:etal:1989}} was used
for the total nucleus-nucleus cross sections. 

For inelastic interactions of hadrons two groups of models are available in the GEANT4 toolkit 
(Agostinelli \etal 2003):\hide{~\cite{Agostinelli:etal:2003}}  
(a) the data-driven models, which are based on the parameterisations of measured 
cross sections for specific reaction channels,   
and (b) the theory-driven models, which are based on various theoretical approaches and implemented
as Monte Carlo event generators.  

In the MCHIT model the inelastic interaction of low-energy (below 20 MeV) nucleons, including radiative
neutron capture, were simulated by means of data driven models. Above 20 MeV the exciton-based precompound
model was invoked (Agostinelli \etal 2003)\hide{~\cite{Agostinelli:etal:2003}}.
 
For hadrons and nuclei with energies above 80A MeV, we have employed
the binary cascade model  (Folger \etal 2004)\hide{~\cite{Folger:etal:2004}}.
In this case exited nuclear remnants are created after the cascade stage of interaction. 
Therefore, appropriate models for the de-excitation process should be included in the simulation.   
The Weisskopf-Ewing model (Weisskopf and Ewing 1940)\hide{~\cite{Weisskopf:and:Ewing:1940}} 
was used for the description of evaporation of nucleons from nuclei at
excitation energies below 3 MeV per nucleon. The Statistical Multifragmentation Model (SMM) by
Bondorf \etal (1995)\hide{~\cite{Bondorf:etal:1995}}
was used at excitation energies above 3 MeV per nucleon
to describe multi-fragment break-up of highly-excited residual nuclei.
The SMM includes as its part the Fermi break-up model
describing an explosive disintegration of highly-excited light nuclei. 

Various unstable nuclei produced in nuclear reactions were followed until
their decay, in particular, the $\beta^+$-decay leading to the emission of a positron.  
Within the GEANT4 toolkit such processes are simulated by using the tables of radioactive 
isotopes with the corresponding decay channels and their probabilities.

\section{Production of  positron-emitting nuclei by protons and carbon ions}

In the case of protons propagating in PMMA  
the positron-emitting nuclei  $^{11}$C, $^{10}$C and $^{15}$O are created
mostly via (p,n) or (p,2n) reactions on $^{12}$C and $^{16}$O nuclei. In this case
all the  positron-emitting nuclei are fragments of target nuclei which are initially at rest. 
In contrast, carbon beams produce $^{11}$C and $^{10}$C nuclei mostly via projectile 
fragmentation, while $^{15}$O nuclei are produced from the target $^{16}$O nuclei.     

The total $\beta^+$-activity yields were measured  for $^{11}$C, $^{10}$C and $^{15}$O 
by Parodi (2004)\hide{~\cite{Parodi:2004}} for protons and carbon ions stopped in PMMA phantoms.
Since the half-lives  of these nuclides are essentially different and well-known from the literature, 
one can build a mathematical model describing their decay and the total rate of 
$\beta^+$-activity as a function of time. The measured time-dependence of the total $\beta^+$-activity
was fitted by Parodi (2004)\hide{~\cite{Parodi:2004}} with the yields of  $^{11}$C, $^{10}$C and $^{15}$O
considered as free parameters. The experimental yields of these nuclides obtained in 
such a way are listed in tables~\ref{EmittersFromProtons} and~\ref{EmittersFromC12} 
along with their uncertainties.

Calculational results of the MCHIT model for $^{11}$C, $^{10}$C and $^{15}$O are listed in 
tables ~\ref{EmittersFromProtons} and~\ref{EmittersFromC12} for comparison.
According to the MCHIT model, the yields of other $\beta^+$-emitting nuclei with $T_{1/2}>1$~min
can be neglected. For example, the yields of $^{13}$N, $^{14}$O, $^{17}$F and $^{18}$F nuclei together 
account for less than 5\% of the total yield of $\beta^+$-emitters produced by $\sim$ 300A~MeV $^{12}$C 
beam in PMMA. The calculations were performed for the PMMA phantoms of the same size 
(300 mm $\times$ 90 mm $\times$ 90 mm) as used by Parodi (2004)\hide{~\cite{Parodi:2004}} in measurements. 
The beam profile in the transverse directions was assumed to be a Gaussian with the FWHM of 10 mm,
while the beam energy spread was taken with the FWHM of 0.2\%.

The results of the
FLUKA and POSGEN codes were reported by Parodi (2004)\hide{~\cite{Parodi:2004}} for several beam energies.
Some of them are listed in tables~\ref{EmittersFromProtons} and~\ref{EmittersFromC12}, respectively.

\begin{table}[htb]
\caption{\label{EmittersFromProtons} Calculated 
yields of  positron-emitting nuclei (per beam particle, in \%\%) produced by 
110, 140 and 175 MeV protons in  PMMA phantom. Experimental data and FLUKA results
by Parodi (2004)\hide{~\cite{Parodi:2004}} are shown for comparison.} 
\begin{indented}
\item[]\begin{tabular}{@{}llllllll}
\br
 & \multicolumn{2}{c}{110 MeV} & \multicolumn{3}{c}{140 MeV} & \multicolumn{2}{c}{175 MeV} \\
\mr
 & MCHIT       & Experiment       & MCHIT           & Experiment          & FLUKA           & MCHIT   & Experiment \\
\mr  
$^{11}$C &  1.83       & $2.2\pm 0.3$&    2.64  & $3.4\pm 0.4   $ &  2.67  & 3.71   & $4.7\pm 0.7$     \\
$^{10}$C &  0.11       & $0.09\pm 0.03$&  0.20  & $0.15\pm 0.03 $ &  0.10  & 0.31   & $0.17\pm 0.06$   \\
$^{15}$O &  0.80       & $0.80\pm 0.15$&  1.10  & $1.23\pm 0.18 $ &  1.23  & 1.54   & $1.6\pm 0.3 $    \\
\br
\end{tabular}
\end{indented}
\end{table}

For protons stopped in the PMMA phantom, the yields of $^{15}$O nuclei predicted by the MCHIT model 
show very good agreement both with the experiment and with the FLUKA results. 
Since $^{11}$C and $^{15}$O are the most abundant and long-lived nuclides, an accurate 
description of their yields is crucial for the PET interpretation. 
As seen from table~\ref{EmittersFromProtons}, the agreement between MCHIT results and the
experiment is within $\sim 20$\% accuracy. As compared with $^{11}$C  and $^{15}$O yields,
the yield of $^{10}$C  is much lower for proton-induced reactions.  The MCHIT results for 
$^{10}$C are in good agreement with the experiment for 110 MeV protons, while for 
140 and 175 MeV protons the theory overestimates $^{10}$C yields 
as compared with the experimental results.      

\begin{table}[htb]
\caption{\label{EmittersFromC12} Calculated
yields of  positron-emitting nuclei (per beam particle, in \%\%) produced by 
212.12A, 259.5A and 343.46A MeV $^{12}$C ions in PMMA phantom.
Experimental data at the same beam energies and POSGEN results (calculations for 270.55A MeV $^{12}$C)
by Parodi (2004)\hide{~\cite{Parodi:2004}} are shown for comparison.
} 
\begin{indented}
\item[]\begin{tabular}{@{}llllllll}
\br
 & \multicolumn{2}{c}{212.12A MeV} & \multicolumn{3}{c}{259.5A MeV} & \multicolumn{2}{c}{343.46A MeV} \\
\mr
 & MCHIT       & Experiment       & MCHIT           & Experiment          & POSGEN           & MCHIT   & Experiment \\
\mr  
$^{11}$C &  11.9       & $10.5\pm 1.3$&   16.83  & $14.7\pm 1.6$ &  26.6     &  25.25  & $19.9\pm 2.4$  \\
$^{10}$C &  1.97       & $ 0.8\pm 0.3$&    2.79  & $1.2\pm 0.3 $ &   1.96    &  4.27   & $1.5\pm 0.3$   \\
$^{15}$O &  2.38       & $ 2.1\pm 0.3$&    3.69  & $3.1\pm 0.4 $ &  10.0     &  6.09   & $5.0\pm 0.4$   \\
\br
\end{tabular}
\end{indented}
\end{table}

As follows from table~\ref{EmittersFromC12}, the predictions of the POSGEN model 
for $^{11}$C and $^{15}$O nuclides, reported by Parodi (2004)\hide{~\cite{Parodi:2004}}
for 270.55A MeV $^{12}$C ions, are less accurate as compared with the MCHIT model in describing
the experimental data at 259.5A MeV. Apparently, more extended benchmarking of the models against each other and 
experimental data is needed to make a firm conclusion about their performance.

The yields of $^{10}$C nuclei are overestimated by the MCHIT model also for the $^{12}$C beam, in particular, 
for the most energetic beam of 343.46A MeV. The POSGEN predictions for the beam energy 270.55A MeV 
are closer to the experimental results obtained at 259.5A MeV.
Nevertheless, we conclude that the MCHIT model is able to describe reasonably well the total yields of 
the most abundant  positron-emitting nuclei, $^{11}$C and $^{15}$O, produced by 
proton and $^{12}$C beams in PMMA.

\section{Spatial distribution of positron-emitting nuclei}

\begin{figure}[htb]  
\begin{centering}
\includegraphics[width=1.00\columnwidth]{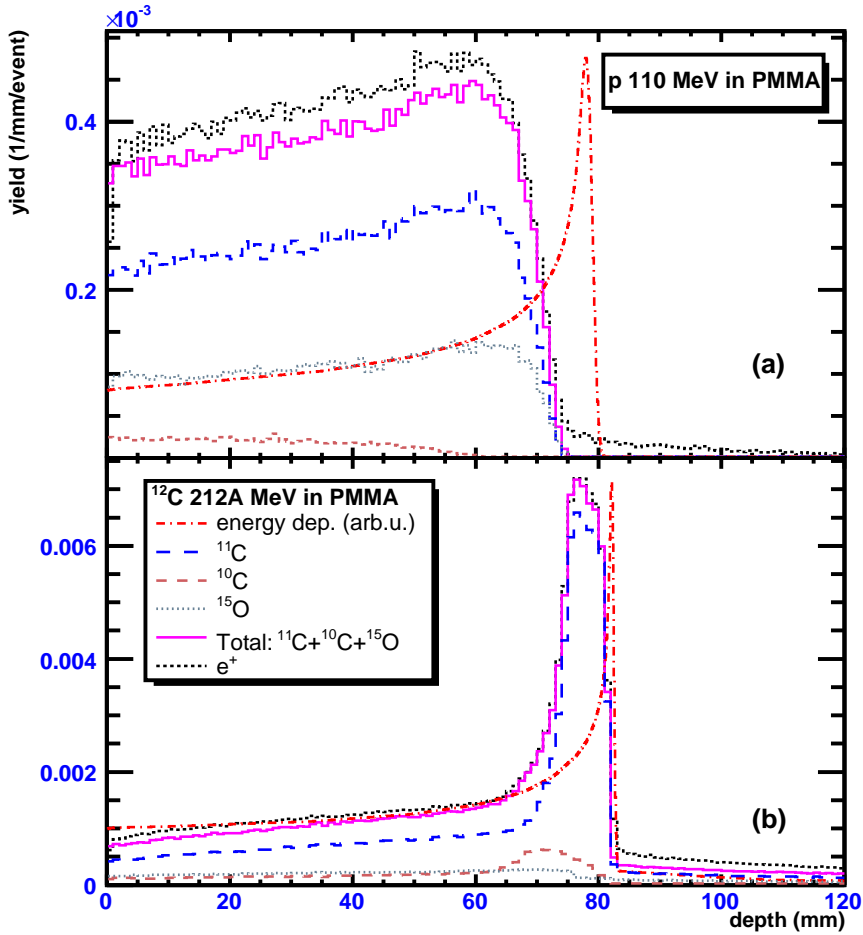}
\end{centering}
\caption{\label{p_C12_NUCLEI} Calculated depth-distributions of deposited energy (dash-dotted curves) and
$\beta^+$-activity (histograms) for (a) 110 MeV protons and  (b) 212.12A MeV $^{12}$C nuclei in PMMA phantom.
The distributions of $^{11}$C, $^{10}$C and $^{15}$O nuclei are shown by the long-dashed, dashed and
dotted histograms, respectively, and their sum is shown by the solid-line histogram.
The distribution of actual e$^+$-annihilation points is shown by the short-dashed histogram.  
}  
\end{figure}

As shown above, the MCHIT model is quite successful in describing the absolute yields of 
$\beta^+$-emitting nuclei. One can now study the spatial distributions of these nuclei.
Such distributions were calculated for 110 MeV protons and 212.12A MeV $^{12}$C ions
together with the corresponding depth-dose distributions for these beams. 
The results are presented in figure~\ref{p_C12_NUCLEI}.  
At these energies, protons and carbon ions have similar ranges in PMMA phantoms.
In both cases the Bragg peaks are located at the depth of $\sim$ 80 mm. 
As demonstrated in the previous section, the contribution from $^{11}$C and $^{15}$O nuclei dominate for both 
the proton and carbon beams. However, the shapes and the absolute values of corresponding $\beta^+$-activity 
distributions are very different.
Protons produce $^{11}$C, $^{10}$C and $^{15}$O fragments from the target carbon and oxygen nuclei.
The cross sections for the $\rm{^{12}C(p,pn)^{11}C}$ and  $\rm{^{16}O(p,pn)^{15}O}$ reactions
increase with decreasing proton energy up to 40-50 MeV and then rapidly fall down.
For proton energies below 20 MeV there is no neutron emission at all. 
As shown in figure~\ref{p_C12_NUCLEI}, $^{11}$C and  $^{15}$O dominate, 
and their yields are slowly increasing as the projectile protons slow down.
The production of $^{10}$C is significantly suppressed because removal of two neutrons without destroying 
the remnant is quite improbable.
As soon as the proton energy becomes below the neutron emission threshold, the 
production of $\beta^+$-activity is ceased, so that no $\beta^+$-emitters are created by protons 
within $\sim 1$ cm before the Bragg peak.  

In contrast to the proton irradiation, the maximum of $\beta^+$-activity produced by carbon ions is
located very close to the Bragg peak and thus clearly marks its position.  
The projectile fragments of interest, $^{11}$C and $^{10}$C, are created 
in peripheral nucleus-nucleus collisions, and have velocities which are very close to the 
projectile velocity. At same initial velocity, the ranges of energetic ions in matter $R$ are proportional to 
$A/Z^2$, where $A$ and $Z$ are the ion mass and charge. Therefore, if $^{11}$C and $^{10}$C are
created at zero depth, their ranges are shorter as compared to the range of projectile 
ions: $R(^{11}{\rm C})\sim 11/12\times R(^{12}{\rm C})$ and $R(^{10}{\rm C})\sim 10/12\times R(^{12}{\rm C})$. 
However, such $^{11}$C and $^{10}$C nuclei can be produced at any depth within the range of primary 
nuclei, excluding only the last few millimetres of the beam range, where $^{12}$C are not energetic 
enough for fragmentation. As discussed in details, in particular, by 
Fiedler \etal (2006)\hide{~\cite{Fiedler:etal:2006}}, the spread in production points leads to the 
spread of stopping points of fragments, as for example, for $^{11}$C fragments this
spread can be estimated as $\Delta x\sim 1/12\times R(^{12}{\rm C})$.  

Despite this spread, all the projectile fragments are stopped before the $^{12}$C Bragg peak.
As shown in figure~\ref{p_C12_NUCLEI}, the sharp fall-off in the $\beta^+$-activity distribution
clearly indicates the position of the Bragg peak.  This is
a very attractive feature of  carbon-ion therapy, which makes possible
on-line monitoring of the ions stopping points. 

The distribution of $^{15}$O nuclei in $^{12}$C induced reactions looks similar to that for the 
proton irradiations, since in both cases $^{15}$O nuclei are produced from the target 
$^{16}$O nuclei. This is why the distribution is quite flat and falls down
before the Bragg peak. Strictly speaking, some of $^{11}$C, $^{10}$C and $^{15}$O nuclei are produced off the
target carbon and oxygen nuclei by secondary particles like p,n and $\alpha$-particles created in 
projectile fragmentation. The contribution of such processes is small, but still visible 
in figure~\ref{p_C12_NUCLEI} beyond the Bragg peak, as light secondary particles propagate further.

From the above analysis we see that $^{11}$C fragments are the most suitable nuclei for monitoring
the energy deposition in carbon-ion therapy. They have the largest yield and longest life time 
($T_{1/2}=20.39$ min) as compared with other $\beta^+$-emitters. Thus, 
mostly $^{11}$C nuclei will survive for 10-20 min after 
stopping, and then emit the positrons.

\section{Distribution of positron annihilation points as measured by a PET scanner}

As explained in section~\ref{Concept}, one can measure the depth distribution of $\beta^+$-activity 
via Positron Emission Tomography (PET). Our present study is focused on the ability of the GEANT4 toolkit 
to describe the production of  positron-emitting nuclei rather then on modelling
various aspects of PET measurements.  As has been already demonstrated by 
Jan \etal (2004),\hide{~\cite{Jan:etal:2004}}
Schmidtlein \etal (2006)\hide{~\cite{Schmidtlein:2006}} and Lamare \etal (2006)\hide{~\cite{Lamare:2006}}, 
the GEANT4 toolkit is very successful in Monte Carlo simulations of commercial PET scanners.

Here we consider only one aspect of the PET monitoring method, i.e. how model predictions 
are affected by a finite resolution of PET scanners. In our model the PET signal is generated in the following 
simplified way. First, the decays of all unstable nuclei are simulated at their stopping points. 
Second, all the emitted positrons are traced up to their annihilation, and, finally, the 
distribution of ${\rm e^+e^-}$-annihilation points is obtained. It is instructive to note
that positrons with energy of a few 100 keV have quite a long mean free path with respect to
annihilation in materials like water, namely $\lambda_{ann}\sim 10$~cm 
($\sigma_{ann}\sim 0.5$~b) at normal water
density. Therefore, the annihilation happens only when the positrons have already slowed down.
 
\begin{figure}[htb]  
\begin{centering}
\includegraphics[width=1.00\columnwidth]{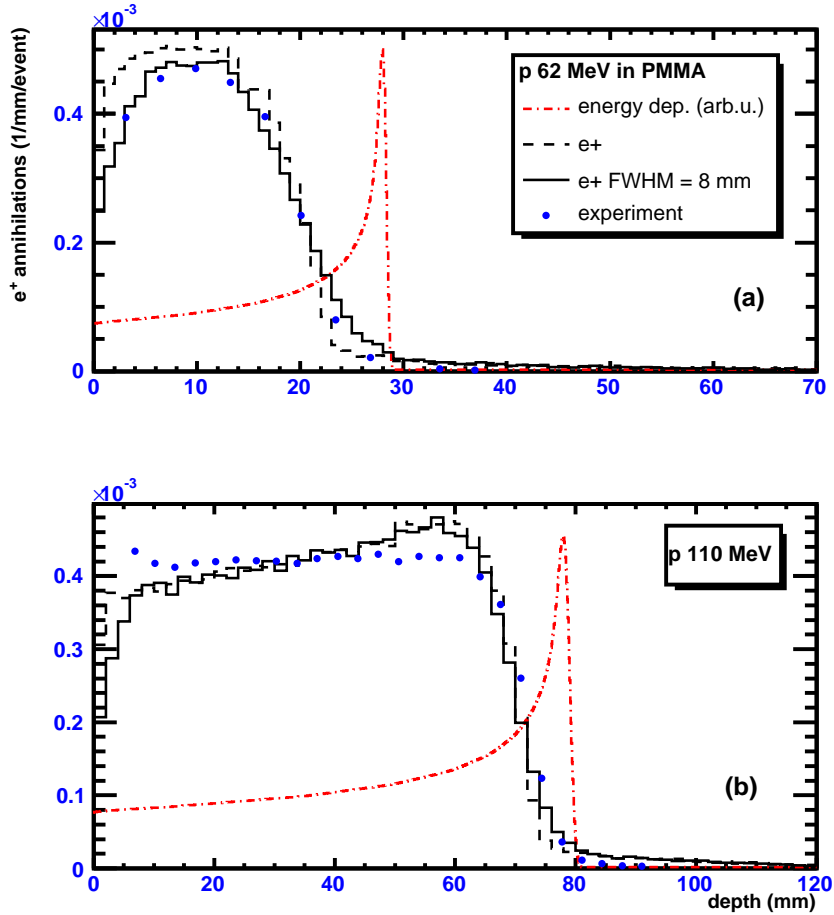}
\end{centering}
\caption{\label{p_lucite} Calculated depth-distributions of deposited energy (dash-dotted curve) and positron
annihilation points (histograms) for (a) 62 MeV and (b) 100 MeV protons in PMMA phantom.  
The distribution of actual e$^+$ annihilation points is shown by the dashed histogram, while
the distribution which accounts for a finite spatial PET resolution 
of FWHM = 8 mm is shown by the solid histogram.
Points show experimental data (Oelfke \etal 1996)\hide{~\cite{Oelfke:etal:1996}}.}
\end{figure}

\begin{figure}[htb]  
\begin{centering}
\includegraphics[width=1.00\columnwidth]{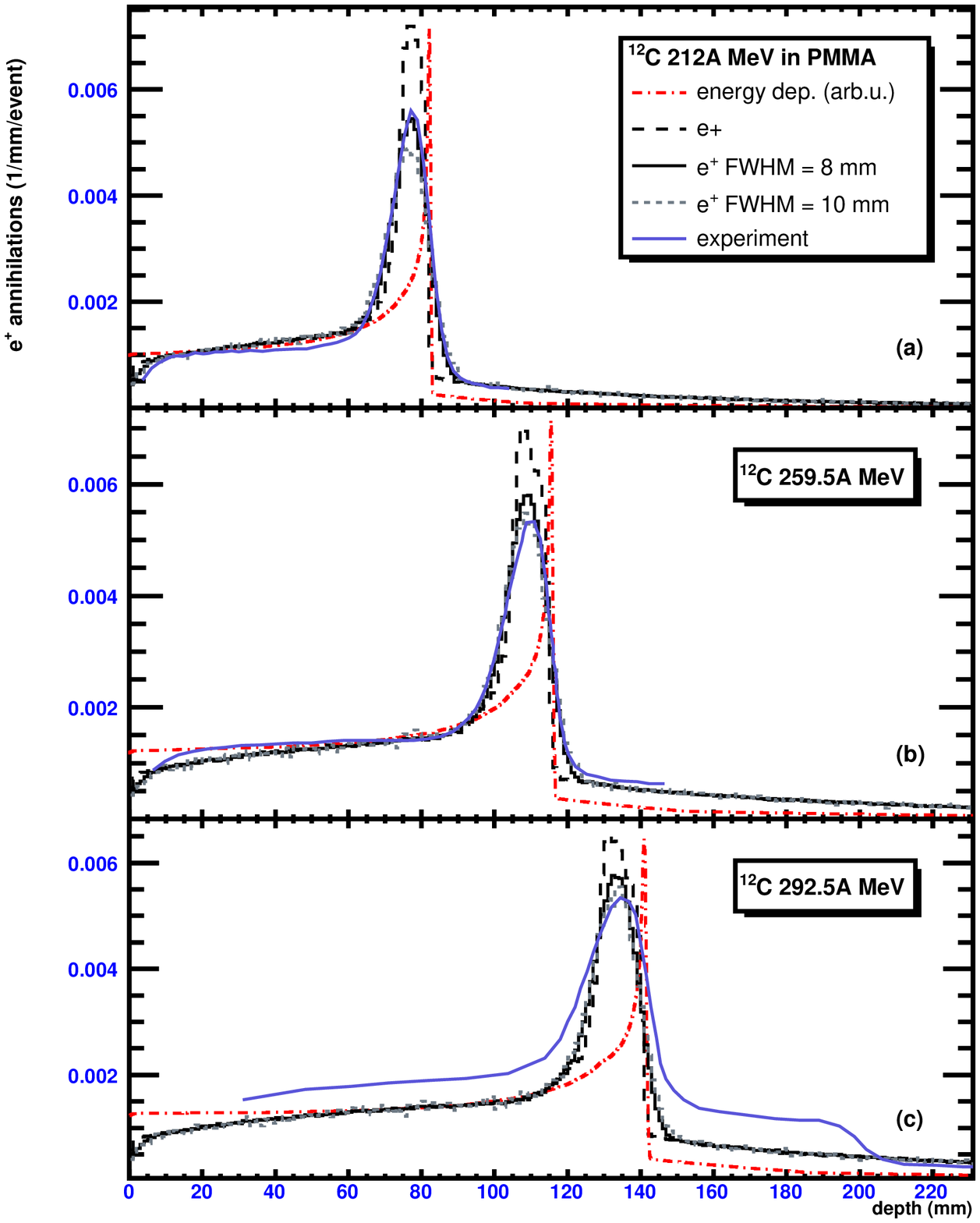}
\end{centering}
\caption{\label{C_12_PMMA_FWHM} Calculated depth-distributions of deposited  energy (dash-dotted curve) and positron
annihilation points (histograms) for $^{12}$C beam with energy 212A MeV (a), 259.5A MeV (b)
and 292.5A MeV (c) 
in PMMA phantom.  The distribution of actual e$^+$ annihilation points is shown by 
the long-dashed histogram, while the distributions which account for a finite spatial PET resolution 
of FWHM = 8 and 10 mm are shown by the  solid and 
short-dashed histograms, respectively.
The experimental data are shown by light solid lines: 
(a)  ( P\"{o}nisch \etal 2004)\hide{~\cite{Poenisch:etal:2004}}  
(b)  (Parodi 2004)\hide{~\cite{Parodi:2004}} and  
(c) (Pawelke \etal 1997)\hide{~\cite{Pawelke:etal:1997}}.  
}
\end{figure}

Third, these distributions are convoluted with Gaussian spreading functions in order to mimic 
a realistic response of a PET imaging system. We choose the FWHM in the range of 8$\div$10 mm.
Such spatial resolution was reported by P\"{o}nisch \etal (2003)\hide{~\cite{Poenisch:etal:2003}} 
for their 3D PET reconstruction algorithm. Results of our simulations for proton and carbon-ion beams 
are shown separately in figures~\ref{p_lucite} and \ref{C_12_PMMA_FWHM}. 
 
Since the measured $\beta^+$-activity distribution depends on the spatial resolution of the PET
scanner, this simple procedure makes possible to compare the calculated  $\beta^+$-activity profiles with those obtained
from experiment, as shown in figures~\ref{p_lucite} and~\ref{C_12_PMMA_FWHM}.  
The distributions of e$^+$ annihilation points generated with the MCHIT model are shown in 
figure~\ref{p_lucite} for 62 and 110 MeV protons in PMMA phantoms. In both cases the distributions
are very wide. When the finite PET resolution is taken into account 
via the convolution of these distributions with a Gaussian of FWHM = 8~mm, 
only small changes in the shape of the distributions are obtained. Nevertheless, this leads to better description of
experimental data by Oelfke \etal (1996)\hide{~\cite{Oelfke:etal:1996}}. The agreement with the experiment
is very good for 62 MeV protons. On the other hand, 
some disagreement with the experimental data is found for 110 MeV protons, i.e. the
measured distribution is more flat as compared to the calculated one.     
In figure~\ref{C_12_PMMA_FWHM} the MCHIT results are compared  with the PET data for  
carbon ions in PMMA phantoms obtained by P\"{o}nisch \etal (2004)\hide{~\cite{Poenisch:etal:2004}} 
for 212.12A MeV, by Parodi (2004)\hide{~\cite{Parodi:2004}} 
for 259.5A MeV and by Pawelke \etal (1997)\hide{~\cite{Pawelke:etal:1997}} for 292.5A MeV.
Some monitoring devices were placed in front of the PMMA phantoms during measurements 
while the MCHIT calculations were performed for pure PMMA phantoms without any additional elements.
To correct for this difference, the experimental data in figure~\ref{C_12_PMMA_FWHM} were shifted by 3 - 6 mm 
to ensure the same position of the Bragg peak in PMMA as reported by Parodi (2004)\hide{~\cite{Parodi:2004}}.
         
For carbon beams, the consideration of the actual PET resolution is crucial for proper description of the
experimental data. In this case the actual distribution of e$^+$ annihilation points consists 
of a sharp peak on a flat plateau. In order to investigate the effect of the PET scanner 
resolution, the calculated distributions
were convoluted with the Gaussians of FWHM = 8 and 10 mm, resulting in the distributions 
shown in figure~\ref{C_12_PMMA_FWHM}. The calculation with FWHM = 8~mm shows a good agreement with 
the experimental data at 212.12A MeV, while for 259.5A MeV a better agreement is achieved for 
FWHM = 10~mm. The MCHIT calculations at 292.5A MeV produce a lower plateau 
as compared with experimental data by Pawelke \etal (1997)\hide{~\cite{Pawelke:etal:1997}}. Since
the measurements in this case were made promptly after irradiations,  the contribution of short-lived isotopes
populating the plateau should be bigger compared to the measurements at 212.12A and 259.5A MeV.
One can note that the calculations by Pawelke \etal (1997)\hide{~\cite{Pawelke:etal:1997}} also
show the peak-to-plateau ratio, which is higher than found in their measurements.

\section{Comparison of dose and $\beta^+$ activity distributions in different materials}

In order to use the PET method for dose monitoring, the employed calculational tool must 
well describe both the dose and $\beta^+$-activity distributions. 
As the $\beta^+$-activity distributions have been already considered above,
we shell study now the corresponding dose distributions calculated with MCHIT, 
which one can then verify
with corresponding experimental data.

\begin{figure}[htb]  
\begin{centering}
\includegraphics[width=1.00\columnwidth]{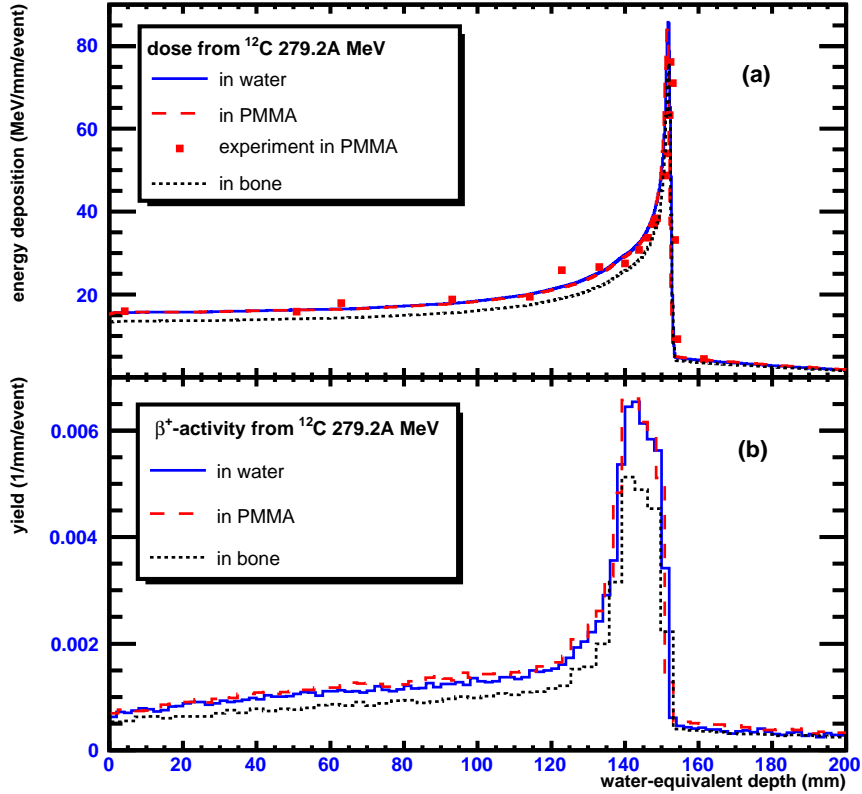}
\end{centering}
\caption{\label{DoseInPMMA} (a) Dose distributions as functions of water-equivalent depth calculated 
for  279.2A MeV $^{12}$C nuclei in water (solid-line histogram), PMMA (dashed histogram) 
and bone (dotted histogram). Experimental data for depth-dose distributions 
in PMMA (Matsufuji \etal 2003)\hide{~\cite{Matsufuji:etal:2003}} are shown by points.
(b) Distributions of positron-emitting nuclei produced in these materials.
}
\end{figure}

The data on dose distributions in PMMA phantoms are rather scarce. 
Matsufuji \etal (2003)\hide{~\cite{Matsufuji:etal:2003}} have performed such measurements  
for beams of $^{4}$He, $^{12}$C, $^{20}$Ne, $^{28}$Si and $^{40}$Ar  nuclei.
In figure~\ref{DoseInPMMA} the calculated dose distribution for 279.2 A MeV $^{12}$C nuclei in PMMA 
is compared with the experimental data of Matsufuji \etal (2003)\hide{~\cite{Matsufuji:etal:2003}}.

As known, the charged particle range in material of a given chemical composition 
is inversely proportional to the density of electrons, and hence, to the matter density $\rho$.     
In particular, it is quite common to present the depth-dose distributions as functions of the 
areal density $x\cdot\rho$. Then, if $\rmd E/\rmd x$ 
is known at one density, it can be easily calculated for other density by rescaling the 
depth $x$.
Strictly speaking, this scaling procedure is fully justified only for a given material at various densities. 
However, one can try to use it also for different tissue-like materials.

In figure~\ref{DoseInPMMA} the scaling property was used to compare the calculated dose and $\beta^+$-activity 
distributions in water with those in PMMA and in dense bone tissue. The elemental composition of 
the bone tissue
was taken in the following mass fractions: H - 6.4\%, C - 27.8\%, N - 2.7\%, O - 41\%, Mg - 0.2\%,
P - 7\%, S - 0.2\%, Ca - 14.7\%. 

In order to get the same position of the Bragg peak, 
the depth in PMMA and bone were stretched by factors of 1.16 and 1.74, respectively, 
which are close to the density ratios, $\rho_{\rm PMMA}/\rho_{\rm water}=1.18$ 
and $\rho_{\rm bone}/\rho_{\rm water}=1.85$. 
In order to account for the changed bin size, the histograms were rescaled by the inverse factors,
$1/1.16$ and $1/1.74$, respectively. Plotted as functions of water-equivalent depth, the dose and $\beta^+$-activity  
profiles in water and PMMA  are in very good agreement with each other 
despite of the difference in the chemical compositions of 
PMMA (${\rm C}_5{\rm H}_8{\rm O}_2$) and water (H$_2$O). However, the calculated position of 
the Bragg peak is $\sim 4$~mm deeper in water compared to the depth-dose profile measured by 
Matsufuji \etal (2003)\hide{~\cite{Matsufuji:etal:2003}}. 
As discussed by Gudowska \etal (2004)\hide{~\cite{Gudowska:etal:2004}}, the  
position of the Bragg peak in experiment was determined with an accuracy of $\pm 3$~mm.
This discrepancy can be attributed to the uncertainty in the initial energy of the 
carbon beam. In order to reach agreement with the experimental distributions the data by
Matsufuji \etal (2003)\hide{~\cite{Matsufuji:etal:2003}} were shifted by 4.2~mm 
towards the larger depth, see figure~\ref{DoseInPMMA}. This is exactly the same shift, which was used
by Gudowska \etal (2004)\hide{~\cite{Gudowska:etal:2004}} to obtain the agreement between the data and 
the results of their SHIELD-HIT code. 

As shown in figure~\ref{DoseInPMMA}, the $\beta^+$-activity distributions in water and PMMA can be successfully 
transformed from one material to another by rescaling. However, this procedure fails for another pair
of materials, water and bone tissue. As one can see in figure~\ref{DoseInPMMA}, the scaling transformation does not
correctly reproduce the Monte Carlo simulation results for the energy deposition in bone tissue. 
The-peak-to-plateau ratio for the $\beta^+$-activity distribution in bone tissue is lower compared to 
water due to the difference in their elemental compositions. As explained above, this peak is
mainly generated by the $^{11}$C nuclei produced from the projectile $^{12}$C nuclei. 
These nuclei are produced less frequently in collisions 
with target nuclei of bone tissue, which are, on average, more heavy compared with water.
In addition, a larger part of the total $\beta^+$-activity ($\sim 10$\%) in bone tissue is due to 
$^{13}$N, $^{14}$O, $^{17}$F, $^{18}$F and $^{30}$P nuclei, which are produced in target fragmentation reactions.
This contribution, which also includes the reactions induced by secondary nucleons beyond the Bragg peak, 
is characterised by a flat depth-distribution. 

We believe that after appropriate validation, our Monte Carlo approach can be used for calculating  
$\beta^+$-activity profiles in various materials, as an alternative method to 
the less accurate scaling transformation. This approach should be especially powerful in the case of 
strongly inhomogeneous medium as human body where soft tissues are intermittent with voids and bones.

\section{Conclusion}

In the present work the validation of the GEANT4 toolkit for calculations in particle therapy 
is extended to the $\beta^+$-activity distributions induced by proton and $^{12}$C beams.
Despite the fact that these  $\beta^+$-activity distributions differ in shape from the corresponding
dose distributions, there exists a strong correlation between them, which can be used for the 
dose and range monitoring by means of the PET scanners.
Since human tissues and organs have various nuclear compositions, only detailed Monte Carlo 
simulations of proton and carbon-ion therapy are able to account for 
the differences in $\beta^+$-activity distributions for different kinds of tissues, e.g. soft tissues and bones 
 
Our MCHIT model based on the GEANT4 toolkit is able to predict with reasonable accuracy
(1) the total yields of $^{11}$C and $^{15}$O nuclei, which are the most abundant $\beta^+$-emitters
produced by proton and $^{12}$C beams, (2) distributions of e$^+$ annihilation points as measured 
by PET scanners with a realistic spatial resolution, (3) the deviations from simple 
density scaling in the dose and $\beta^+$-activity distributions.
 By this study, the ability of the GEANT4 
toolkit to correctly simulate the dose and $\beta^+$-activity distributions in the proton and carbon 
ion therapy has been clearly demonstrated.
Therefore, we suggest to use the MCHIT model for realistic calculations of $\beta^+$-activity profiles and 
dose distributions in proton and carbon-ion therapy.

\ack

This work was partly supported by Siemens Medical Solutions.
We are grateful to Prof. Hermann Requardt for the discussions which stimulated the present study. 
The discussions with Dr. Thomas Haberer, Prof. J\"{u}rgen Debus and Dr. Katia Parodi are gratefully acknowledged.


\References 

\item[] Agostinelli S \etal  (GEANT4 Collaboration) 2003 GEANT4: 
A simulation toolkit {\it Nucl. Instrum. Meth. A } {\bf 506} 250-303

\item[]   Amaldi U  2004
 CNAO - The Italian Centre for Light-Ion Therapy
{\it Radiother. Oncol.} {\bf 73}  S191-201

\item[]  Bajard M, De Conto J M and Remillieux J 2004
Status of the "ETOILE" project for a French hadrontherapy centre   
{\it Radiother. Oncol.} {\bf 73} S211-5

\item[] Bennett G W, Goldberg A C, Levine G S, Guthy J, Balsamo J and Archambeau J O  1975
Beam localization via O-15 activation in proton-radiation therapy
{\it Nucl. Instr. Meth.} {\bf 125} 333-8

\item[] Bennett G W, Archambeau J O, Archambeau B E, Meltzer J I and Wingate C L  1978 
Visualization and transport of positron emission from proton activation in vivo 
{\it Science} {\bf 200} 1151-3

\item[] Bondorf J P, Botvina A S, Iljinov A S, Mishustin I N and Sneppen K  1995 
Statistical multifragmentation of nuclei {\it Phys. Rept.} {\bf 257} 133-221

\item[] Carrier J F, Archambault L, Beaulieu L and Roy R 2004
 Validation of GEANT4, an object-oriented Monte Carlo toolkit, for simulations in medical physics
{\it Med. Phys.} {\bf 31} 484-92

\item[] Castro J R, Petti P L, Blakely E A and Daftari I K 2004
Particle radiation therapy 
{\it Textbook of Radiation Oncology} (Saunders, Elsevier Inc.)
ed Leibel S A and Phillips T L 
pp 1547-68 

\item[] Enger S A, Rezaei A, af Rosenschold P M and Lundqvist H 2006
Gadolinium neutron capture brachytherapy (GdNCB), 
a new treatment method for intravascular brachytherapy
{\it Med. Phys.} {\bf 33} 46-51

\item[] Enghardt W, Fromm W D, Geissel H, Keller H, Kraft G, Magel A, Manfrass P, 
M\"{u}nzenberg G, Nickel F, Pawelke J, Schardt D, Scheidenberger C and Sobiella M  1992
The spatial-distribution of positron-emitting nuclei generated by relativistic light-ion
beams in organic-matter
{\it Phys. Med. Biol.} {\bf 37}  2127-31

\item[] Fern\'{a}ndez-Varea J M, Andreo P and Tabata T  1996
Detour factors in water and plastic phantoms and their use for range and depth scaling 
in electron-beam dosimetry
{\it Phys. Med. Biol.} {\bf 41} 1119-39

\item[] Fiedler F, Crespo P, Parodi K, Sellesk M, Enghardt W 2006
The feasibility of in-beam PET for therapeutic beams of $^{3}$He
{\it IEEE Trans. Nucl. Sci.} {\bf 53} 2252-9

\item[] Folger  G, Ivanchenko V N and Wellisch J P  2004 The Binary Cascade - 
nucleon-nuclear reactions {\it Eur. Phys. J. A} {\bf 21} 407-17

\item[] GEANT4-Documents 2006 {\it http://geant4.web.cern.ch/geant4/G4UsersDocuments/Overview/html/}

\item[] GEANT4-Webpage 2006 {\it http://geant4.web.cern.ch/geant4/}

\item[]  Griesmayer E and Auberger T  2004 
The status of MedAustron
{\it Radiother. Oncol.} {\bf 73} S202-5

\item[] Gudowska I, Sobolevsky N, Andreo P, Belkic D and Brahme A  2004 
Ion beam transport in tissue-like media using the Monte Carlo code 
SHIELD-HIT {\it Phys. Med. Biol.} {\bf 49} 1933-58

\item[]  Haberer T, Debus J, Eickhoff H, J\"{a}kel O, Schulz-Ertner D and Weber U 2004
The Heidelberg ion therapy center 
{\it Radiother. Oncol.} {\bf 73} S186-90

\item[]  Heeg P, Eickhoff H and Haberer T 2004
Conception of heavy ion beam therapy at Heidelberg University (HICAT)
{\it Z. Med. Phys.}  {\bf 14} 17-24

\item[]  Hishikawa Y, Oda Y, Mayahara H, Kawaguchi A, Kagawa K, Murakami M and Abe M 2004
Status of the clinical work at Hyogo
{\it Radiother. Oncol.} {\bf 73}  S38-40

\item[] Hong L, Goitein M, Bucciolini M, Comiskey R, Gottschalk B, Rosenthal S, 
Serago C and Urie M  1996 A pencil beam algorithm for proton dose calculations 
{\it Phys. Med. Biol.} {\bf 41} 1305-30

\item[]  Iseki Y, Kanai T, Kanazawa M, Kitagawa A, Mizuno H, Tomitani T, Suda M and Urakabe E 2004 
Range verification system using positron emitting beams for heavy-ion radiotherapy
{\it Phys. Med. Biol.} {\bf 49} 3179-95

\item[] J\"{a}kel O, Kr\"{a}mer M, Karger C P and Debus J  2001 
Treatment planning for heavy ion radiotherapy: clinical implementation 
and application {\it Phys. Med. Biol.} {\bf 46} 1101-16

\item[] Jan S, Santin G, Strul D \etal 2004 
GATE: a simulation toolkit for PET and SPECT
{\it Phys. Med. Biol.} {\bf 49} 4543-61

\item[]  Jiang H and Paganetti H  2004
Adaptation of GEANT4 to Monte Carlo dose calculations based on CT data
{\it Med. Phys.} {\bf 31} 2811-8

\item[]  Kanai T, Matsufuji N, Miyamoto T, Mizoe J, Kamada T, Tsuji H, Kato H, Baba M and Tsujii H 2006
Examination of GyE system for HIMAC carbon therapy
{\it Int. J. Radiat. Oncol. Biol. Phys.} {\bf 64} 650-6

\item[] Kraft G 2000
Tumor therapy with heavy charged particles 
{\it Prog. Part. Nucl. Phys.} {\bf 45} S473-544

\item[] Kr\"{a}mer M, J\"{a}kel O, Haberer T, Kraft G, Schardt D and Weber U  2000 
Treatment planning for heavy-ion radiotherapy: physical beam model and dose 
optimization {\it Phys. Med. Biol.} {\bf 45} 3299-317

\item[] Lamare F, Turzo A, Bizais Y, Rest C C and Visvikis D 2006 
Validation of a Monte Carlo simulation of the Philips Allegro/GEMINI 
PET systems using GATE
{\it Phys. Med. Biol.} {\bf 51} 943-62 

\item[]  Larsson S, Svensson R, Gudowska I, Ivanchenko V and Brahme A  2005
 Radiation transport calculations for 50 MV photon therapy beam using the Monte Carlo code GEANT4
{\it Radiat. Prot. Dosimetry.} {\bf 115} 503-7

\item[] Levin C S and Hoffman E J  1999
Calculation of positron range and its effect on the fundamental limit of positron emission 
tomography system spatial resolution
{\it Phys. Med. Biol.} {\bf 44} 781-99

\item[] Matsufuji  N, Fukumura A, Komori M, Kanai T and Kohno T 2003
Influence of fragment reaction of relativistic heavy charged particles on heavy-ion radiotherapy
{\it Phys. Med. Biol.} {\bf  48} 1605-23

\item[] Mayahara H, Oda Y, Kawaguchi A, Kagawa K, Murakami M, Hishikawa Y, 
Igaki H, Tokuuye K and Abe M  2005
A case of hepatocellular carcinoma initially treated by carbon ions, 
followed by protons for marginal recurrence with portal thrombus
{\it Radiat. Med.} {\bf 23} 513-9

\item[]  Oelfke U, Lam G K and Atkins M S  1996
Proton dose monitoring with PET: quantitative studies in Lucite
{\it Phys. Med. Biol.} {\bf 41} 177-96

\item[] Parodi K and Enghardt W  2000
Potential application of PET in quality assurance of proton therapy
{\it Phys. Med. Biol.} {\bf 45} N151-6

\item[] Parodi K, Enghardt W and Haberer T  2002
In-beam PET measurements of $\beta^+$ radioactivity induced by proton beams
{\it Phys. Med. Biol.} {\bf 47} 21-36

\item[] Parodi K, 2004
On the feasibility of dose quantification with in-beam PET data in radioterapy 
with $^{12}$C and proton
beams 2004, Ph.D. Dissertation, Technische Universit\"{a}t Dresden

\item[] Perez-Calatayud J, Granero D, Ballester F, Crispin V and Van der Laarse R 2006
Technique for routine output verification of Leipzig applicators with a well chamber
{\it Med. Phys.} {\bf 33} 16-20

\item[] Pawelke J, Byars L, Enghardt W, Fromm W D, Geissel H, Hasch B G, Lauckner K, 
Manfrass P, Schardt D and Sobiella M  1996
The investigation of different cameras for in-beam PET imaging
{\it Phys. Med. Biol.} {\bf 41} 279-96

\item[] Pawelke J, Enghardt W, Haberer T, Hasch B G, Hinz R, Kramer M, 
Lauckner K and Sobiella M   1997
In-beam PET imaging for the control of heavy-ion tumour therapy
{\it IEEE Trans. Nucl. Sci.} {\bf 44} 1492-8

\item[]   P\"{o}nisch F, Enghardt W and Lauckner K 2003
Attenuation and scatter correction for in-beam positron emission tomography 
monitoring of tumour irradiations with heavy ions
{\it Phys. Med. Biol.} {\bf 48} 2419-36

\item[] P\"{o}nisch F, Parodi K, Hasch B G and Enghardt W 2004
The modelling of positron emitter production and PET imaging during carbon ion therapy 
{\it Phys. Med. Biol.} {\bf 49} 5217-32

\item[] Pshenichnov I, Mishustin I and Greiner W 2005
Neutrons from fragmentation of light nuclei in tissue-like media: a study with the GEANT4 toolkit
{\it Phys. Med. Biol.} {\bf 50} 5493-507

\item[] PTCOG 2006 Particle Therapy Cooperative Group Webpage
http://ptcog.web.psi.ch/

\item[] Rogers D W O 2006 
Fifty years of Monte Carlo simulations for medical physics  
{\it Phys. Med. Biol.} {\bf 51} R287-301

\item[] Schmidtlein C R, Kirov A S, Nehmeh S A, Erdi Y E, Humm J L, Amols H I, 
Bidaut L M, Ganin A, Stearns C W, McDaniel D L and Hamacher K A  2006
Validation of GATE Monte Carlo simulations of the GE Advance/Discovery 
LS PET scanners
{\it Med. Phys.} {\bf 33} 198-208

\item[] Schulz-Ertner D, Nikoghosyan A, Thilmann C, Haberer T, J\"{a}kel O, Karger C, Kraft G, 
Wannenmacher M and Debus J 2004
Results of carbon ion radiotherapy in 152 patients
{\it Int. J. Radiat. Oncol. Biol. Phys.} {\bf 58} 631-40

\item[] Shen W, Wang B, Feng J, Zhan W, Zhu Y and Feng E 1989 
Total reaction cross-section for heavy-ion collisions and its relation to the neutron 
excess degree of freedom
{\it Nucl. Phys.} {\bf A491} 130-46 

\item[]  Tobias C A, Benton E V, Capp M P, Chatterjee A, Cruty M R and Henke R P 1977
Particle radiography and autoactivation
{\it Int. J. Radiat. Oncol. Biol. Phys.} {\bf 3} 35-44

\item[]  Tsujii H, Mizoe J E, Kamada T, Baba M, Kato S, Kato H, Tsuji H, 
Yamada S, Yasuda S, Ohno T, Yanagi T, Hasegawa A, Sugawara T, Ezawa H, 
Kandatsu S, Yoshikawa K, Kishimoto R and Miyamoto T 2004
Overview of clinical experiences on carbon ion radiotherapy at NIRS.
{\it Radiother. Oncol.} {\bf 73} S41-9

\item[] Urakabe E, Kanai T, Kanazawa M, Kitagawa A, Noda K, Tomitani T, Suda M, Iseki Y, 
Hanawa K, Sato K, Shimbo M, Mizuno H, Hirata Y, Futami Y, Iwashita Y and Noda A  2001
Spot scanning using radioactive C-11 beams for heavy-ion radiotherapy
{\it Jpn. J. Appl. Phys.} {\bf 40} 2540-8

\item[] Weisskopf V E and Ewing D H  1940  On the yield of nuclear reactions with 
heavy elements {\it Phys. Rev.} {\bf 57} 472-85

\item[] Wellisch H P and Axen D 1996 Total reaction cross section calculations in proton-nucleus scattering
{\it Phys. Rev. C} {\bf 54} 1329-32

\endrefs

\end{document}